\pdfoutput=1

\documentclass[11pt,twoside,a4paper,cmspaper,final,collab]{cms-tdr}
\def\svnVersion{250614}\def\cmsCernNo{2013-223}\def\cmsCernDate{\today}\def\cmsMessage{Published in Physics Letters B as \href{http://dx.doi.org/10.1016/j.physletb.2014.06.041}{\doi{10.1016/j.physletb.2014.06.041.}}}

\begin{document}\cmsNoteHeader{SMP-12-026}

\hyphenation{had-ron-i-za-tion}
\hyphenation{cal-or-i-me-ter}
\hyphenation{de-vices}

\RCS$Revision: 246287 $
\RCS$HeadURL: svn+ssh://svn.cern.ch/reps/tdr2/papers/SMP-12-026/trunk/SMP-12-026.tex $
\RCS$Id: SMP-12-026.tex 246287 2014-06-13 15:49:34Z alverson $
\newcolumntype{x}{D{,}{\pm}{-1}}
\newlength\cmsFigWidth
\ifthenelse{\boolean{cms@external}}{\setlength\cmsFigWidth{0.45\textwidth}}{\setlength\cmsFigWidth{0.48\textwidth}}
\ifthenelse{\boolean{cms@external}}{\providecommand{\cmsLeft}{top}}{\providecommand{\cmsLeft}{left}}
\ifthenelse{\boolean{cms@external}}{\providecommand{\cmsRight}{bottom}}{\providecommand{\cmsRight}{right}}
\newcommand{\ZZ}{\cPZ\cPZ\xspace}
\newcommand{\WW}{\PWp\PWm\xspace}
\newcommand{\WZ}{\PW\cPZ\xspace}
\newcommand{\pp}{\Pp\Pp\xspace}
\newcommand{\Wjets}{\ensuremath{\PW+\text{jets}}\xspace}
\newcommand\Wmnbb{\ensuremath{\PW(\Pgm\cPgn)+\bbbar}}
\newcommand\WmnH{\ensuremath{\PW(\Pgm\cPgn)\PH}}
\newcommand\Wbb{\ensuremath{\PW+\bbbar}\xspace}
\newcommand\Wc{\ensuremath{\PW+\cPqc}}
\newcommand\Wmn{\ensuremath{\PW\to \Pgm\Pgn}}
\newcommand\Wln{\ensuremath{\PW\to \ell\Pgn}}
\newcommand\Wcc{\ensuremath{\PW+\ccbar}}
\newcommand\Zjets{\ensuremath{\cPZ+\text{jets}}}
\newcommand\Zll{\ensuremath{\cPZ\to\ell\ell}}
\newcommand\Zmm{\ensuremath{\cPZ\to \Pgm\Pgm}}
\newcommand\MT{M_\mathrm{T}}
\newcommand{\Lint}{\ensuremath{\mathcal{L}_{\text{int}}}}
\newcommand{\dytt}{\ensuremath{\cPZ/\Pgg^*\to\tau^+\tau^-}}
\newcommand{\dyll}{\ensuremath{\cPZ/\Pgg^*\to \ell^+\ell^-}}
\newcommand{\DPSerr}{\ensuremath{\,\mathrm{(DPS)}}\xspace}
\newcommand{\haderr}{\ensuremath{\,\text{(had.)}}\xspace}
\newcommand{\MCFMerr}{\ensuremath{\,\mathrm{(MCFM)}}\xspace}

\providecommand{\theo}{\ensuremath{\,\text{(theo.)}}\xspace}
\providecommand{\vecPtmu}{\ensuremath{\ptvec(\mu)}\xspace}

\cmsNoteHeader{SMP-12-026}
\title{Measurement of the production cross section for a W boson and two b jets in pp collisions at \texorpdfstring{$\sqrt{s} = 7\TeV$}{sqrt(s)=7 TeV}}

\date{\today}
\abstract{The production cross section for a $\PW$ boson and two b jets is measured using proton-proton collisions at $\sqrt{s} = 7\TeV$ in a data sample collected with the CMS experiment at the LHC corresponding to an integrated luminosity of  5.0\fbinv. The $\PW+\bbbar$ events are selected in the $\PW\to\Pgm\Pgn$ decay mode by requiring a muon with transverse momentum $\pt>25\GeV$ and pseudorapidity $\abs{\eta}<2.1$, and exactly two b-tagged jets with $\pt>25\GeV$ and $\abs{\eta}<2.4$. The measured $\PW+\bbbar$ production cross section in the fiducial region, calculated at the level of final-state particles, is $\sigma(\Pp\Pp\to \PW+\bbbar) \times {\mathcal{B}}(\PW\to\mu\nu)=0.53\pm 0.05\stat \pm 0.09\syst \pm 0.06\theo \pm 0.01\lum\unit{pb}$, in agreement with the standard model prediction. In addition, kinematic distributions of the $\PW+\bbbar$ system are in agreement with the predictions of a simulation using \MADGRAPH and \PYTHIA.}
\hypersetup{%
pdfauthor={CMS Collaboration},%
pdftitle={Measurement of the production cross section for a W boson and two b jets in pp collisions at sqrt(s) = 7 TeV},
pdfsubject={SMP},%
pdfkeywords={CMS, physics, SMP}
}
\maketitle 

\section{Introduction}
This Letter reports a study of the production of a $\PW$ boson and two b jets in proton-proton collisions,
where the $\PW$ boson is observed via its decay to a muon and a neutrino,
and each b jet is identified by the presence of a b hadron with a displaced decay vertex.
The production mechanism of $\bbbar$ pairs together with
$\PW$ or $\cPZ$ bosons has been the subject of extensive
theoretical studies and is included
in different simulation programs~\cite{Oleari:2011ey,Frederix:2011qg,Madgraph5}, but is still not thoroughly understood.
Previous
measurements of vector boson production with associated b-quark jets
have shown varying levels of agreement with theoretical
calculations~\cite{Aaltonen:2009qi,D0:2012qt,Aad:2011kp}.

According to the standard model (SM), the primary contribution for $\bbbar$ production in association with a W boson
is due to the splitting of a gluon into a $\bbbar$ pair.
Two different models for b-quark production are available, depending on whether there are four or five quark flavors
in the proton parton distribution functions (PDFs)~\cite{Compare4F5F}.
Therefore, a precise experimental measurement of the $\Wbb$ production cross section
provides important input to the refinement of theoretical calculations in
perturbative quantum chromodynamics (QCD), as well as the validation of
Monte Carlo (MC) techniques.

A key feature of this analysis compared to others~\cite{Aaltonen:2009qi,D0:2012qt,Aad:2011kp}
is the $\bbbar$ phase space that is covered.
Previous measurements
have concentrated on $\PW$-boson production with at least
one observed b-quark jet, for which the predictions differ from the experimental results.
This difference is larger in the production of events with a collinear $\bbbar$ pair
that is reconstructed as one jet~\cite{ANGULARCORRELATIONSZBB,HiggsBBCMS},
a topology afflicted by significant theoretical uncertainties.
Focusing on the observation of $\PW$-boson production with
two well-separated b-quark jets, this analysis provides a
complementary approach by
probing a kinematic regime that is better understood theoretically.

The production of $\Wbb$ events is an irreducible background
in analyses involving two
separated and well-identified b jets,
such as SM Higgs boson production in association with
an electroweak gauge boson and subsequent decay to
$\bbbar$.
The discovery of a Higgs boson with a mass of approximately 125\GeV
by the ATLAS and CMS Collaborations~\cite{HiggsAtlas,HiggsCMS,LongHiggsCMS} motivates
further studies to determine the coupling of this new boson to b quarks.

Other SM processes produce events with an experimental signature similar to the one studied here. These include
production of top quark-antiquark pairs ($\ttbar$), associated production of a
$\PW$ boson with light jets misidentified as b-quark jets, single-top-quark
production, multijet production (henceforth labeled ``QCD multijet''), Drell--Yan production associated with jets, and electroweak diboson production.

\section{CMS detector and event samples}

This analysis uses a sample of proton-proton  collisions
at a center-of-mass energy of $\sqrt{s}=7\TeV$, collected in 2011 with the Compact Muon Solenoid (CMS)
experiment at the LHC, and corresponding to an
integrated luminosity  $\int{ \textrm{L} \, \rd{}t}$ of 5.0\fbinv.
While the CMS detector is described in detail elsewhere~\cite{CMSExperiment}, the
key components for this analysis are summarized below.
The CMS experiment uses a right-handed coordinate system, with the
origin at the nominal interaction point, the $x$ axis
pointing to the center of the LHC ring, the $y$ axis pointing
up (perpendicular to the plane of the LHC ring), and the $z$ axis
along the counterclockwise-beam direction. The polar angle
$\theta$  is measured from the positive $z$ axis and the
azimuthal angle  $\phi$  is measured in the $x$-$y$ plane in radians.
The magnitude of the transverse
momentum $\PT$ is calculated as
$\PT = \sqrt{\smash[b]{p_{x}^2 + p_{y}^2}}$.
A superconducting solenoid is the central feature
of the CMS detector, providing an axial magnetic
field of 3.8\unit{T} parallel to the beam direction.
A silicon pixel and strip
tracker, a crystal electromagnetic calorimeter, and a brass/scintillator hadron
calorimeter are located within the solenoid. A quartz-fiber
Cherenkov calorimeter extends the coverage to $\abs{\eta} <5.0$, where  $\eta=-\ln[\tan{(\theta/2)}]$.
Muons are measured in gas-ionization detectors embedded
in the steel flux return yoke outside the solenoid.
The first level of the CMS trigger system, composed of custom
hardware processors, is designed to select the most interesting events
using information from the calorimeters and muon
detectors.
A high-level trigger processor farm decreases the event rate
to a few hundred hertz, before data storage.

A number of MC event generators are used to simulate the signal and
background event samples.
Vector boson + jets and $\ttbar$ + jets production
are generated at leading order (LO)
using \MADGRAPH~5.1~\cite{Madgraph5} interfaced with
\PYTHIA~6.4.24~\cite{Sjostrand:2006za}
for hadronization. The $\Wjets$ sample was generated using the five-flavor scheme,
which includes massless b quarks in the initial state.
Single-top-quark event samples are generated at next-to-leading order (NLO) with
\POWHEG~2.0~\cite{Alioli:2008gx,Nason:2004rx,Frixione:2007vw}.
Diboson ($\WW$, $\WZ$, $\ZZ$) samples are
generated with
\PYTHIA~6.4.24.
For LO generators, the default PDF set used is CTEQ6L~\cite{CTEQ66}, while for NLO generators the
showering of partons and hadronization are simulated with \PYTHIA using the Z2
tune~\cite{Field:2010bc}.
For all processes, the detector response is simulated using a detailed
description of the CMS detector based on  \GEANTfour~\cite{GEANT}.
The reconstruction of simulated events is performed with
the same algorithms used for the analyzed data sample.
The simulated event samples include additional minimum-bias interactions per bunch crossing
(pileup).

\section{Event reconstruction}

Individual particles emerging from each collision are reconstructed with the
particle-flow (PF) technique~\cite{CMS-PAS-PFT-09-001, CMS-PAS-PFT-10-002}. This
approach uses the information from all subdetectors to identify and
reconstruct individual particle candidates in the event, classifying
them into mutually exclusive categories: charged hadrons, neutral hadrons, photons, electrons, and muons.

Muons are reconstructed
by combining the information from the tracker and the muon
spectrometer~\cite{CMS-PAS-MUO-10-002}.
The muon candidates are required to originate from the primary vertex of the
event, chosen as the vertex with the highest $\sum \pt^2$ of the charged particles associated with it.
The muon relative isolation is defined as 
$I^{\rm{rel}} = \sum_{i}\pt(i)/\pt(\mu)$,
with $i$ running over PF candidates (hadrons, electrons, photons)
in a cone around the muon direction defined by $\Delta R < 0.4$, where $\Delta R = \sqrt {\smash[b]{ (\Delta \eta)^2 + (\Delta \phi)^2}}$.

The $\ptvec$ of a muon passing the identification and isolation requirements is
combined with the missing transverse energy $\VEtmiss$ of the event to form a $\PW$ candidate.
We define $\VEtmiss$ as
the negative vector sum of the
transverse momenta of all reconstructed particle candidates in the event.
The value of $\VEtmiss$ is corrected for noise in the electromagnetic and hadron calorimeters using the procedure described in Ref.~\cite{WZCMS:2010}, which is
based on a parametrization of the recoil energy measured in $\mathrm{Z}\to\mu\mu$ events.
The reconstructed transverse mass of the system, $\MT$, is calculated from the
$\pt$ of the isolated muon and the $\VEtmiss$ of the event;
$\MT = \sqrt{\smash[b]{2\, \pt(\mu)\, \abs{\VEtmiss}\, (1-\cos \Delta \phi)}}$,
where
$\Delta \phi$
is the difference in azimuth between $\VEtmiss$ and $\vecPtmu$. In $\Wmn$ decays,
the $\MT$ distribution exhibits a Jacobian peak with a kinematic
endpoint at the $\PW$ mass. It is therefore a natural
discriminator against non-$\PW$ final states, such as QCD multijet events, that have a lepton candidate and $\VEtmiss$ and a relatively low value of $\MT$.

Jets are constructed using the anti-\kt clustering algorithm~\cite{Cacciari:2008gp}, as implemented in the \textsc{fastjet} package~\cite{fastjet1,fastjet2},
with a distance parameter of 0.5. Jet clustering is performed using individual particle candidates reconstructed with the PF technique.
Jets are required to pass identification
criteria that eliminate jets originating from
noisy channels in the hadron calorimeter~\cite{Chatrchyan:2009hy}.
Jets originating from pileup interactions are
rejected by requiring consistency of the jets
with the primary interaction vertex.
Small corrections to jet energy---relative and absolute calibrations of the detector---are
applied as a function of the $\pt$ and $\eta$  of the jet~\cite{cmsJEC}.

The combined secondary vertex (CSV) b-tagging algorithm~\cite{BTAGNOTE} exploits
the long lifetime and relatively large mass of b hadrons to provide optimized
b-quark jet discrimination.
The CSV algorithm combines information about impact parameter
significance, secondary vertex (SV) kinematic properties, and jet kinematic
properties in a likelihood-ratio technique.
Jets are b-tagged by imposing a minimum threshold on the CSV discriminator value.
This threshold provides an efficiency of approximately 50\% for identifying
jets containing b-flavored hadrons, while limiting the misidentification
probability for light-quark and gluon jets to 0.1\% and for c-quark jets to 3\%.
Furthermore, to increase the purity of the sample,
a selected jet is required to have a reconstructed SV.
This additional requirement has
a small impact on the selected b-quark jets (93$\%$
efficiency with respect to the b-tag selection) while reducing the
combined misidentification probability for c-quark, light-quark, and gluon jets to ${<}$0.1\%.

\section{\texorpdfstring{$\Wbb$}{W to b b-bar} event selection \label{sectionEventSelection}}

Candidate events are selected online by a single-muon trigger
that requires a reconstructed muon with $\pt>24\GeV$ and $\abs{\eta}<2.1$.
The offline $\Wbb$ event selection requires
an isolated muon with $I^{\text{rel}}<0.12$, $\pt>25\GeV$, $\abs{\eta}<2.1$, and
exactly two jets with $\pt>25\GeV$ and $\abs{\eta}<2.4$, where
both selected jets must contain a SV
and pass the b-tagging requirement.
To reduce the contribution from $\cPZ$-boson production, the
event is rejected if a second muon forms an
invariant mass $m_{\mu\mu} > 60\GeV$ with the isolated muon. There are no requirements
on the isolation or $\pt$ of the second muon.
The $\ttbar$ background is reduced by requiring
that there be no additional isolated electrons or muons with  $\pt>20\GeV$
in the event and no additional jets with $\pt>25\GeV$ and $2.4<\abs{\eta}<4.5$. To reduce the
contribution from QCD multijet events, $\MT > 45\GeV$ is required.
The total number of observed events in the data sample after all selection requirements are applied is 1230.

We first estimate the normalized distributions for the signal and
each type of background and then
perform a global fit to determine the fraction of each background
in the candidate sample.
The shapes of the signal and background distributions for the variables we use in the fit are evaluated using simulation,
except for the QCD multijet background, which is derived from data.
The cross sections for the  $\Wjets$ and $\Zjets$ processes are calculated with the predictions from  \textsc{fewz}~\cite{fewz}
evaluated at next-to-next-to-leading order (NNLO) using the
MSTW08~NNLO PDF set~\cite{Martin:2009iq}.
Single-top-quark and diboson production cross sections are
normalized to the NLO
cross section predictions from $\MCFM$~\cite{Campbell:2010ff, Badger:2010mg}
using the MSTW08~NLO PDF set. The $\ttbar$ cross section is
taken at NNLO as calculated in Ref.~\cite{TOPNNLO}.
For each background simulation we apply the same selection requirements
as for the
candidate sample to generate the relevant distributions for fitting.

To prove that the simulation describes the data both in shape and normalization, and
can be used in the fit as described in the following section,
we select specific data samples
that are enriched with the relevant backgrounds
and verify the performance of the simulation. These control
regions are not used in the final signal extraction and serve only for
this verification task, with the exception of the QCD multijet control region.

The shapes of the distributions for multijet events are taken directly from a
multijet enriched  data sample
obtained using the signal selection requirements and, in addition, requiring a nonisolated muon: $I^{\text{rel}}>0.2$.
The yield of multijet events is obtained from a fit performed on
 events with $M_{T} < 40\GeV$, which is below the Jacobian peak of the $\Wmn$.
The resulting
normalization is extrapolated to the signal region, $\MT>45\GeV$.
The relative uncertainty in the yield of QCD multijet events is
estimated to be $\pm$50\%, taking into account both the fit result and the extrapolation to the high-$\MT$ range. This relative
uncertainty also covers
shape mismodelings of the small multijet contribution in the final sample.

The  $\PW+\text{light-quark}$ jets  process, where the jets are not initiated by b or c quarks,
is the dominant background before applying the selection requirements on the SV  and
on b-tagging.
The b-tagging algorithm reduces the contamination of light-quark and c-quark jets in the selected sample to approximately 2\% of the total expected yield.
The contribution of events with a single b-quark jet in the initial state and a misidentified second light-quark or gluon jet is negligible.

\begin{figure}[htbp]
\centering
\includegraphics[width=\cmsFigWidth]{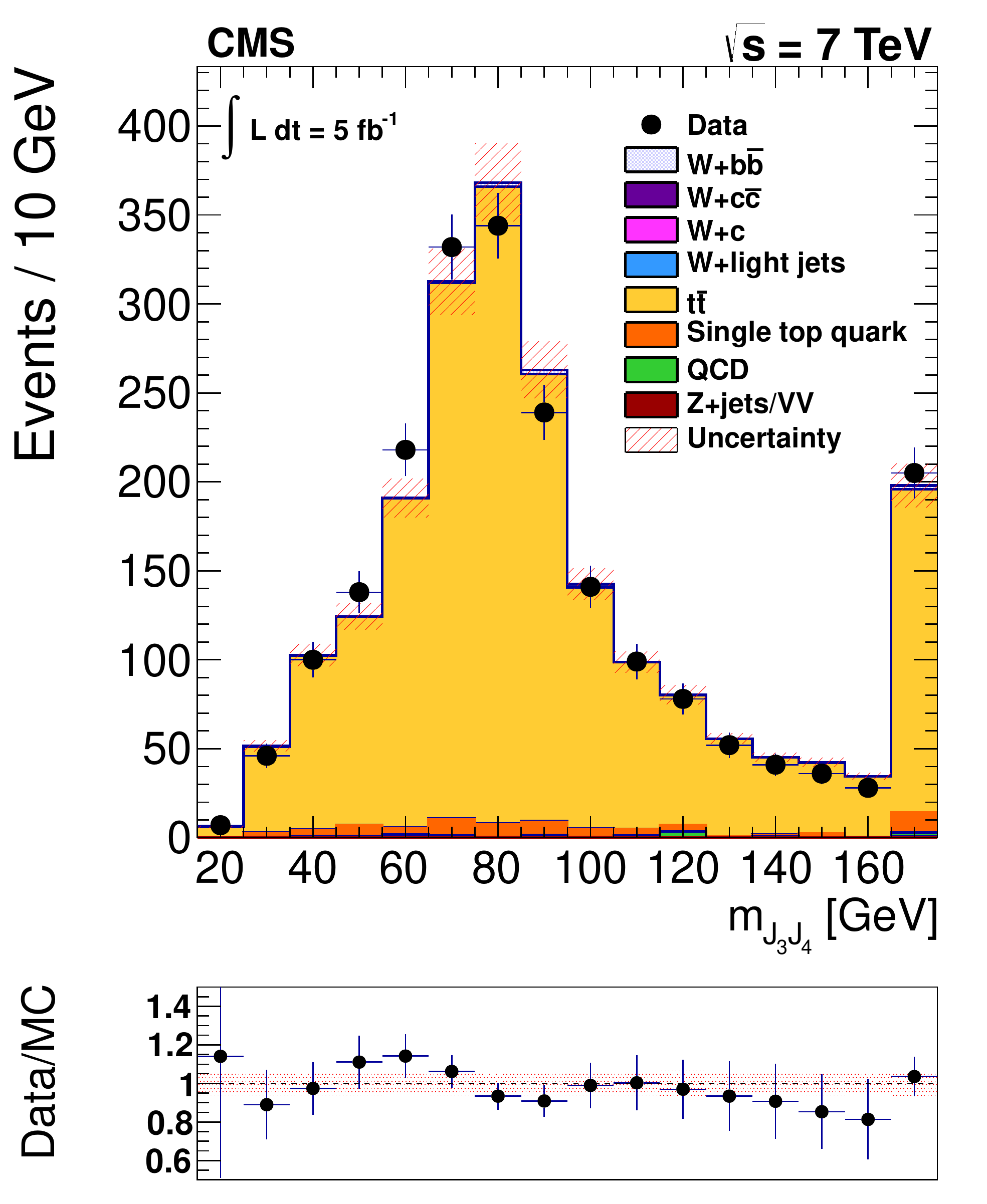}
\includegraphics[width=\cmsFigWidth]{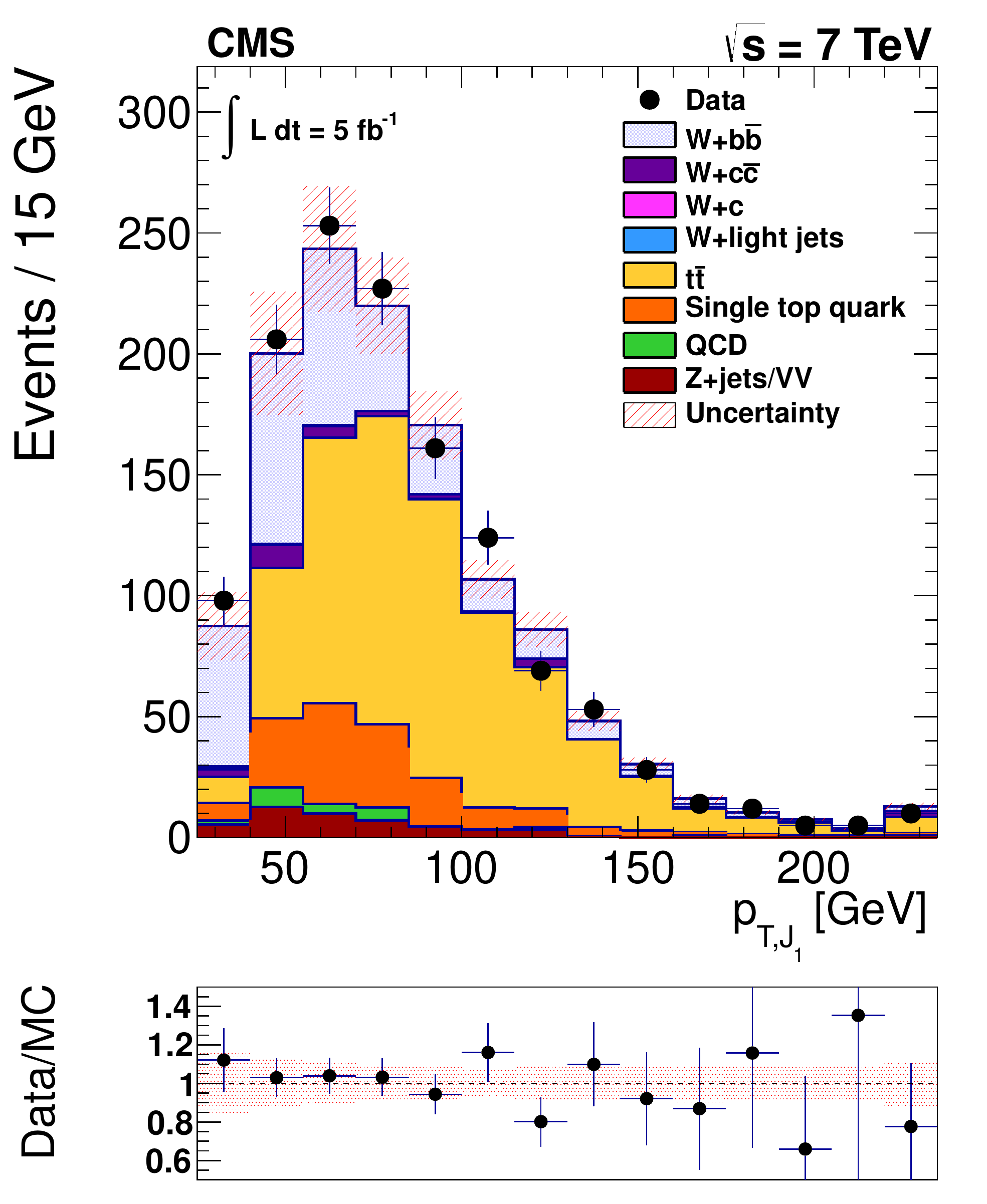}
\caption{
The distribution of the invariant mass $m_{\mathrm{J_3 J_4}}$ of the two additional light jets
in the $\ttbar$ control region (\cmsLeft).
The $\pt$ distribution of the highest-$\pt$ jet, $p_{\mathrm{T},\mathrm{J_1}}$, in the signal region (\cmsRight).
Signal and background yields are taken from the maximum-likelihood fit to CMS data described in the text.
The uncertainty band corresponds to the total uncertainty on the fitted yields.
The last bin in both figures includes overflow events. The lower panels show the ratio of observed data events
to the total fitted yield. }
\label{figB}
\end{figure}

A $\ttbar$ background control data sample is formed by requiring
two jets in addition to the two highest $\pt$ b-tagged jets.
This higher jet multiplicity requirement selects a sample that is dominated by $\ttbar$ events.
Figure~\ref{figB} (\cmsLeft) shows the invariant mass $m_{\mathrm{J_{3}J_{4}}}$ of
the third- and fourth-highest $\pt$ jets in the event.
In $\ttbar$ events this observable is correlated to the mass of the
hadronically decaying $\PW$ boson.
This $\ttbar$ control region is used in the final fit
for the signal yield
to constrain the $\ttbar$ background normalization in the signal region.
The simulation describes the observed distributions well, both in
terms of shape and normalization.

A $\Zjets$ background data sample is defined by requiring the standard selection
criteria with the additional requirement of a second muon with opposite charge
such that the invariant mass of the dimuon system is consistent
with a $\cPZ$ boson ($70 < m_{\mu\mu} < 100\GeV$).
This sample is used to validate the $\Zjets$ background estimate, as documented in Ref.~\cite{CMSZBB}.
The simulation describes  the experimental distributions well in this control data sample.

A single-top-quark background sample is defined by selecting events in which the
$\PW$ boson is accompanied by exactly one b-quark jet,
which passes the tagging criteria, and an additional forward jet with $\abs{\eta}>2.8$.
No further rejection of additional light jets or leptons is imposed.
The simulation describes  the single-top-quark background data
sample well, as documented in  Ref.~\cite{CMSSINGLETOPTCHANNEL}, and therefore
it is used to estimate the yield and shape of the distributions of kinematic variables in the signal region.

The expected yield in the signal region for the SM Higgs boson of $M_\PH=125\GeV$
associated with a $\PW$ boson
where the Higgs boson decays to $\bbbar$ pairs and the $\PW$ boson decays to a muon and a neutrino
has been computed using the
\POWHEG event generator. It would account
for ${<}$0.2\% of the total expected yield in the signal region
and is not considered for this measurement.

\section{Signal extraction}

After all the selection requirements, the largest background
contributions are the production of $\ttbar$ pairs and single top quarks.
Contributions to the background from $\Wjets$, $\Zjets$, diboson production, and QCD
multijet events are much smaller, as shown in the middle column of Table~\ref{tab:fitYields}.

The composition of the candidate data sample is extracted
via a binned extended maximum-likelihood fit.
Because the $\ttbar$ background is large (larger in fact than the signal), it is essential to constrain tightly both
the signal and the $\ttbar$ background with a simultaneous fit
to the $\pt$ of the
leading jet ($p_{\mathrm{T},\mathrm{J_1}}$) in the signal region after all selection requirements are applied,
and to the $m_{\mathrm{J_3 J_4}}$ distribution obtained from the $\ttbar$ control sample.
The normalization of each background contribution is allowed to vary in the fit within its uncertainty.
The  $\pt$ of the leading jet is chosen as the final fit variable because of its
discrimination power against top-related backgrounds.
Figure~\ref{figB} shows the fitted distributions: $p_{\mathrm{T},\mathrm{J_1}}$ in the
signal region (\cmsRight) and $m_{\mathrm{J_3 J_4}}$ in the $\ttbar$ control sample
(\cmsLeft);  the yields shown for the different processes are those resulting from
the fit. The $\chi^2$ of the fit is  16.9, for 29 degrees of freedom ($\chi^2/\mathrm{dof}=0.58$).
The fitted yields for all  the processes are listed in Table~\ref{tab:fitYields},
and  compared to the predictions. All observed yields are found to be in agreement
with the expectations.

The systematic uncertainties, including those in the predicted background yields,
are introduced as nuisance parameters in the fit with constraints around the estimated central value. Any cross section
or acceptance uncertainty in the background processes is introduced as a log-normal constraint on the rate of the process.
Alternate binned templates are obtained by varying the different sources of systematic uncertainty; the nominal
and alternate templates are then interpolated depending on the nuisance
parameter values.
One of the largest systematic uncertainties comes from the
relative uncertainty in the b-tagging efficiency (6\% per jet). This and other
uncertainties in the light-quark and c-quark jet mistagging efficiencies are
taken from Ref.~\cite{BTAGNOTE}.
The jet energy and muon $\pt$ scales are allowed to vary within their uncertainties (1--3\%  and 0.2\%, respectively).
Relative uncertainties in the muon efficiency (due to trigger, reconstruction, identification, and isolation) are estimated to be 1\%.
The average number of  pileup events in the data sample analyzed is 9.
The uncertainty associated with the pileup in the simulation is studied by shifting the overall mean
number of interactions per bunch crossing up or down by 0.6, which has a negligible effect on the measurement.
To account for the uncertainty in the description of the $\VEtmiss$ spectrum, the component of $\VEtmiss$ that is not
clustered in jets is shifted by ${\pm}$10\%.
The normalizations of the background processes are also taken into account, with an uncertainty assigned to each process
according to the theoretical predictions, the previous CMS measurements when available,
or an estimate from the multijet data sample.
The overall relative uncertainty in the signal selection efficiency due to the choice of PDF set
is estimated by following the PDF4LHC recommendation
and found to be approximately 1\%~\cite{Alekhin:2011sk,Botje:2011sn,Lai:2010vv,Martin:2009iq,Ball:2011mu}.
The varying of the factorization ($\mu_{\mathrm{F}}$) and renormalization ($\mu_{\mathrm{R}}$) scales,
also based on the {PDF4LHC} recommendation, leads to an uncertainty of 10\%.
A similar procedure is followed to estimate the effect of scale variations
on the signal shape, yielding an uncertainty in the cross section smaller than 1\%.
The relative integrated luminosity uncertainty is 2.2\%~\cite{CMS-PAS-SMP-12-008}.

\begin{table}[htb]
\begin{center}
\topcaption{Comparison of the yields expected from the SM
and obtained from the fit to the analyzed data sample. The predicted yield uncertainty
takes into account the uncertainties in the measurement of the cross section for each of
the processes, except for the multijet contribution, which
is estimated using a multijet enriched background data sample.
The uncertainty in the fitted yields combines the statistical and systematic uncertainties
obtained from the extended binned maximum-likelihood fit technique.
}
\label{tab:fitYields}
\begin{tabular}{c x x }
\hline
Process  & \multicolumn{1}{c}{Predicted yield}  & \multicolumn{1}{c}{Fitted yield} \\
\hline
$\Wbb$         & 332,99     & 301,59 \\
$\ttbar$       & 621,36     & 653,37 \\
Single top quark   & 160,13     & 167,13 \\
QCD multijet   & 33,17      & 33,16  \\
$\Wc$, $\Wcc$  & 21,4       & 20,10 \\
$\Zjets$         & 31,3       & 32,3 \\
WW, WZ         & 19,3       & 19,3   \\
$\PW+\text{light-quark}$ jets         & 1.5,0.2    & 1,1  \\
\hline
Total          & 1219,79    & 1226,73 \\
\hline
Observed events & \multicolumn{2}{c}{$1230$} \\
\hline
\end{tabular}
\end{center}
\end{table}

\begin{figure}[htbp]
\centering
\includegraphics[width=\cmsFigWidth]{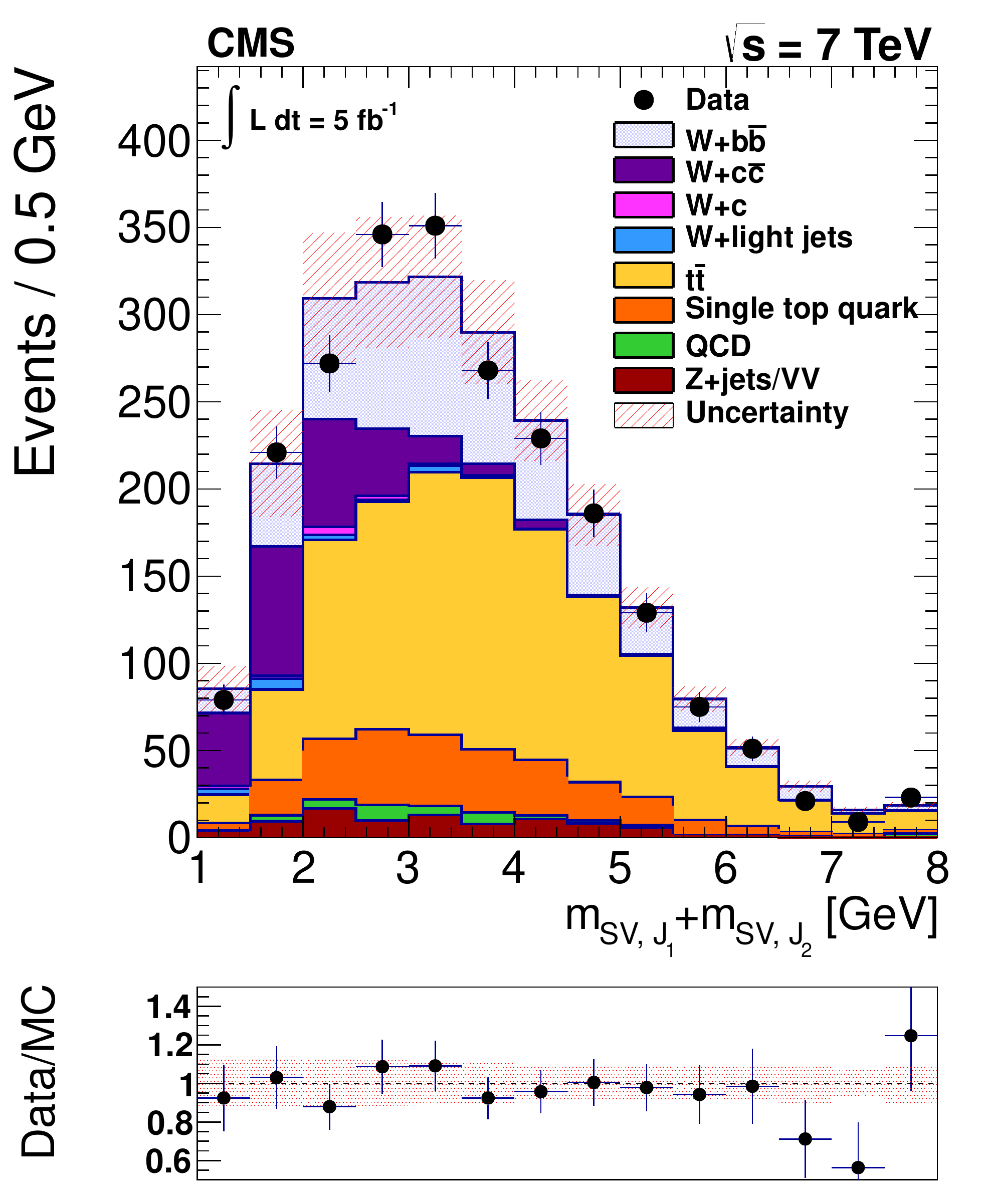}
\includegraphics[width=\cmsFigWidth]{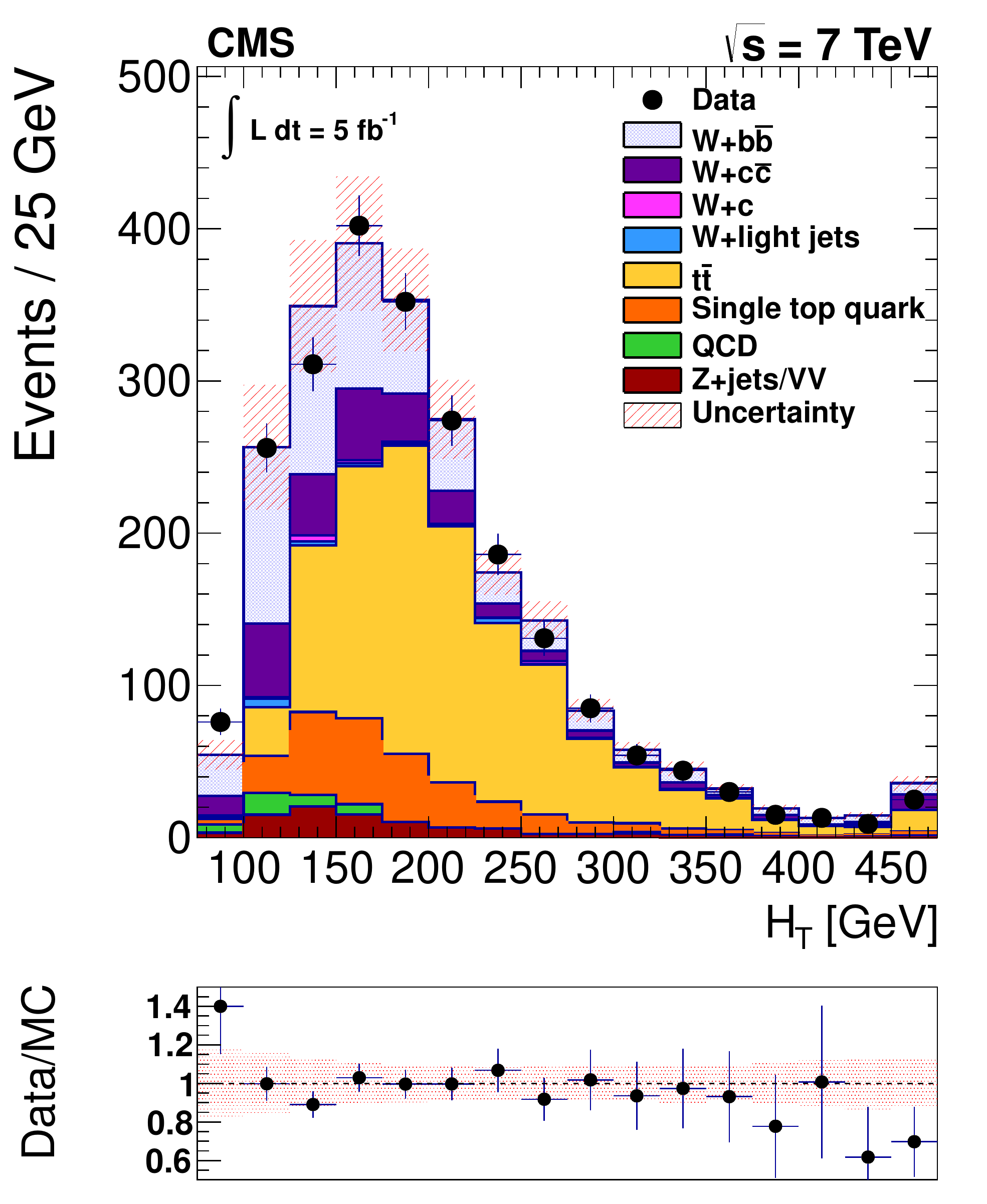}
\caption{
The distribution of the sum of the masses reconstructed from all the particles originating at the
SVs, $m_{\mathrm{SV},\,\mathrm{J_1}}+m_{\mathrm{SV},\,\mathrm{J_2}}$, (\cmsLeft) and the distribution for the variable $H_{\mathrm{T}}$
(\cmsRight) in the alternative looser b-tag selection. These two distributions are the
projections of the two-dimensional fit performed as a cross-check and described in the text.
Signal and background yields correspond to the post-fit results. The uncertainty band corresponds
to the total uncertainty in the fitted yields. The last bin in both figures includes overflow events.
The lower panels show the ratio of observed data events
to the total fitted yield.
}
\label{figC}
\end{figure}

The number of events in the candidate sample, 1230, is in agreement with the expected and fitted total yields, although it is not explicitly included in the fitting process.
The number of signal events obtained from the binned maximum-likelihood fit is $N_{S} = 301 \pm 30\stat \pm 51\syst$.

To cross-check these results, an independent study is performed with looser b-tagging criteria, corresponding
to an efficiency  of 70\% for selecting a jet containing b-flavored hadrons,
while the misidentification probability for light-quark and gluon jets is 1\% and for c-quark jets is 11\%.
All other selection criteria for the signal
and control samples remain unchanged.
Since the c-quark jet contribution becomes significant with these looser criteria,
it is essential to use variables in the fit that can discriminate against both
$\Wcc$ and top-quark-initiated processes.
The invariant mass measured using all particles originating at the SVs
of the highest $\pt$ ($m_{\mathrm{SV},\,\mathrm{J_1}}$) and second-highest $\pt$ ($m_{\mathrm{SV},\,\mathrm{J_2}}$) jets
can distinguish between $\Wbb$ and $\Wcc$. The scalar sum of the transverse momenta
of the jets,
$H_{\mathrm{T}}$, is used to distinguish $\Wjets$ from
top-quark contributions.
The $\Wbb$ signal is extracted in a two-dimensional fit using
the two variables $m_{\mathrm{SV},\,\mathrm{J_1}}+m_{\mathrm{SV},\,\mathrm{J_2}}$ and $H_{\mathrm{T}}$ and
constraining the $\ttbar$ contribution in
the $\ttbar$ background data sample as described above.
The distributions of the variables $m_{\mathrm{SV},\,\mathrm{J_1}}+m_{\mathrm{SV},\,\mathrm{J_2}}$
and $H_{\mathrm{T}}$, which are projections of the two-dimensional distributions
fitted in this cross-check,  are shown in Fig.~\ref{figC}, with yields
as given by the fit.
The central value of the cross section computed with this method differs by less than 3\% from
the primary fit result.

\section{Results}

The $\Wbb$ cross section can be measured within a fiducial volume defined by requiring
a final-state muon with $\pt>25$\GeV and $\abs{\eta}<2.1$ and exactly two final-state particle jets, reconstructed using the anti-\kt
jet algorithm with a distance parameter of 0.5, with $\pt>25$\GeV and $\abs{\eta}<2.4$ and
with each containing at least one b hadron with $\pt>5$\GeV. Events with extra jets with $\pt>25$\GeV and $\abs{\eta}<4.5$ are vetoed.

Within this fiducial phase space, the $\Wbb$ cross section
is obtained using the expression
\begin{equation*}
\sigma(\Pp\Pp\to \Wbb) \times {\mathcal{B}}(\Wmn)= \frac{N_{S}}{\int{\mathrm{L}\,\rd{}t }~\epsilon_{\text{sel}}},
\end{equation*}
where the efficiency of the selection requirements, $\epsilon_{\text{sel}} = (11.2 \pm 1.0)\%$, is computed using the
\MADGRAPH + \PYTHIA MC sample. The uncertainty in this selection efficiency comes from the PDF
and scale variation uncertainties mentioned above. The experimental uncertainties are included in the determination of $N_{S}$.

The measured fiducial cross section  is
\ifthenelse{\boolean{cms@external}}{
\begin{multline*}
\sigma(\Pp\Pp\to \PW + \bbbar) \times {\mathcal{B}}(\Wmn) =0.53\pm  0.05\stat \pm\\
 0.09\syst \pm 0.06\theo \pm 0.01\lum\unit{pb.}
\end{multline*}
}{
\begin{equation*}
\sigma(\Pp\Pp\to \PW + \bbbar) \times {\mathcal{B}}(\Wmn) =0.53\pm  0.05\stat \pm 0.09\syst \pm 0.06\theo \pm 0.01\lum\unit{pb.}
\end{equation*}
}

This measured value cannot be directly compared to the SM NLO cross section
calculated with \MCFM~\cite{Campbell:2010ff, Badger:2010mg} because the latter
pertains to jets of partons, not jets of hadrons, and does not include
the production of $\bbbar$ pairs from double-parton scattering (DPS).

\MCFM predicts a cross section of $0.52 \pm 0.03\unit{pb}$ at the parton level,
using the {MSTW2008} NNLO PDF set and setting the factorization
and renormalization scales to $\mu_{\mathrm{F}} = \mu_{\mathrm{R}} = m_{\PW} + 2m_{\cPqb}$~\cite{Badger:2010mg}.
The uncertainty in the theoretical cross section quoted above
is estimated
by varying the scales $\mu_{\mathrm{F}}$, $\mu_{\mathrm{R}}$ simultaneously
up and down by a factor of two. It also takes into account the
PDF uncertainties following the PDF4LHC recommendation.
The scale uncertainty in the theoretical cross section may be underestimated because of
the requirement of exactly two jets in the final state, which introduces a veto on events with
extra jets. Therefore, a more conservative
estimate of this uncertainty in the theoretical prediction is computed, following the procedure
described in Ref.~\cite{JetUncertainty}, and the total theoretical uncertainty is found to be 30\%.

Two corrections are needed to link the
theoretical prediction to the  measurement, a hadronization correction and a DPS correction.
At the parton level, the events are required to have a muon of
$\pt > 25\GeV$ and $\abs{\eta}<2.1$ and exactly two parton jets of
$\pt > 25\GeV$ and $\abs{\eta}<2.4$, each containing a b quark.
The hadronization correction factor $C_{\cPqb \to \PB}=0.92\pm0.01$, calculated using a five-flavor
\MADGRAPH + \PYTHIA reference MC, is used to extrapolate the cross section computed
at the level of parton jets to the level of final-state particle jets.
The uncertainty assigned to this correction
is obtained by comparing the corresponding factors computed with a four-flavored \MADGRAPH MC simulation.
The simulated \MADGRAPH + \PYTHIA events include DPS
production of $\bbbar$ pairs and they reproduce these processes adequately
as measured by CMS~\cite{CMS_DPS}. The contribution of DPS events to the cross section
at the parton-jet level is estimated to be $\sigma_{\mathrm{DPS}} = (\sigma_{\PW} \times \sigma_{\bbbar})/\sigma_{\text{eff}}=0.08\pm0.05\unit{pb}$.
The value of the effective cross section, $\sigma_{\text{eff}}$, is taken from Ref.~\cite{ATLAS_DPS},
and is assumed to be independent of the process and interaction scale.
The uncertainty in $\sigma_{\mathrm{DPS}}$ takes into account both the uncertainty in the measurement of $\sigma_{\text{eff}}$
and the uncertainty in the fiducial $\bbbar$ cross section.
The theoretical cross section at hadron level can be extrapolated from the MCFM parton-jet prediction by applying the hadronization correction and
adding the DPS contribution,
resulting in $0.55 \pm0.03\MCFMerr \pm 0.01\haderr  \pm 0.05\DPSerr\unit{pb}$.
This value is in agreement with the measured value.

\begin{figure}[htb]
\centering
\includegraphics[width=\cmsFigWidth]{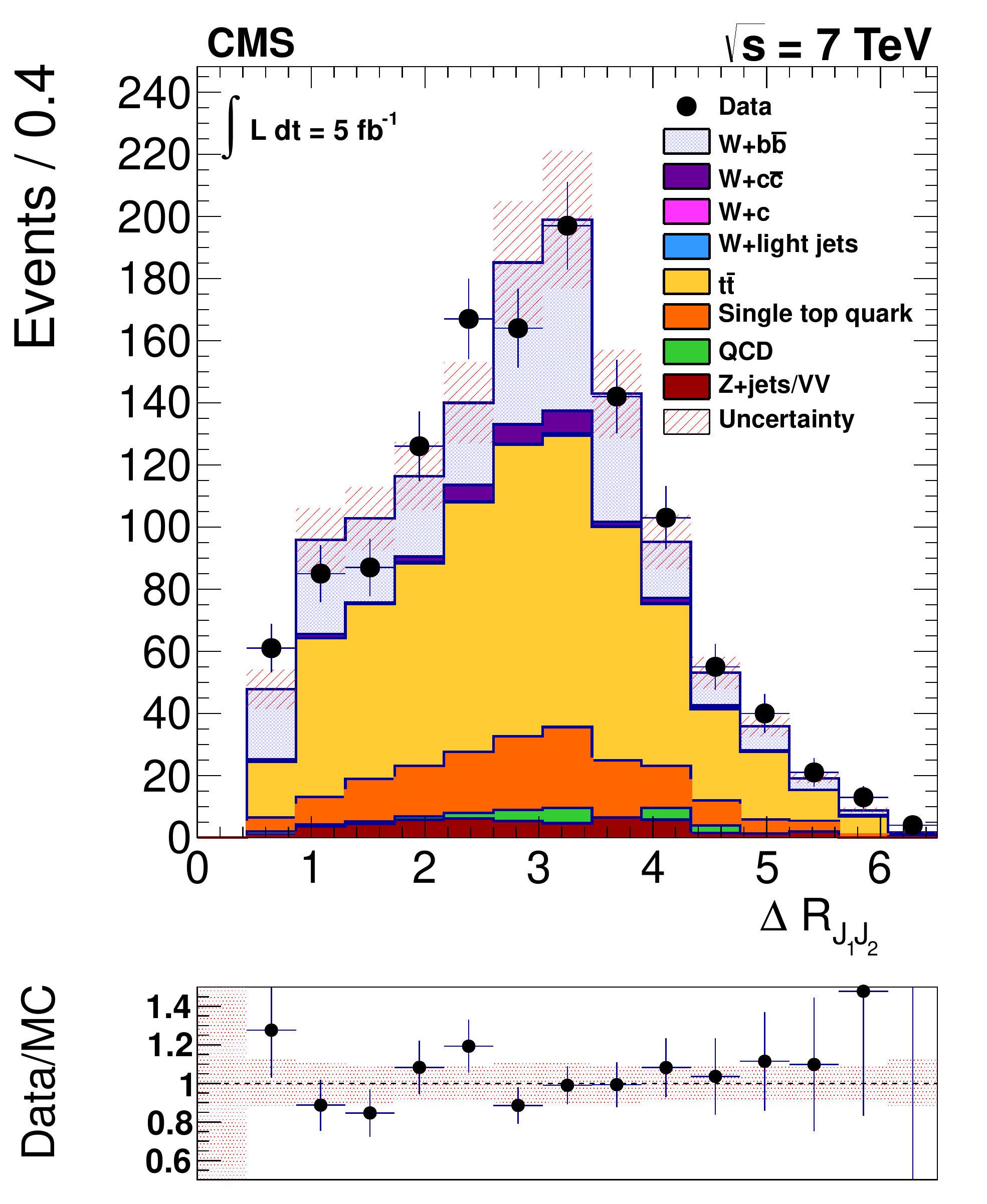}
\includegraphics[width=\cmsFigWidth]{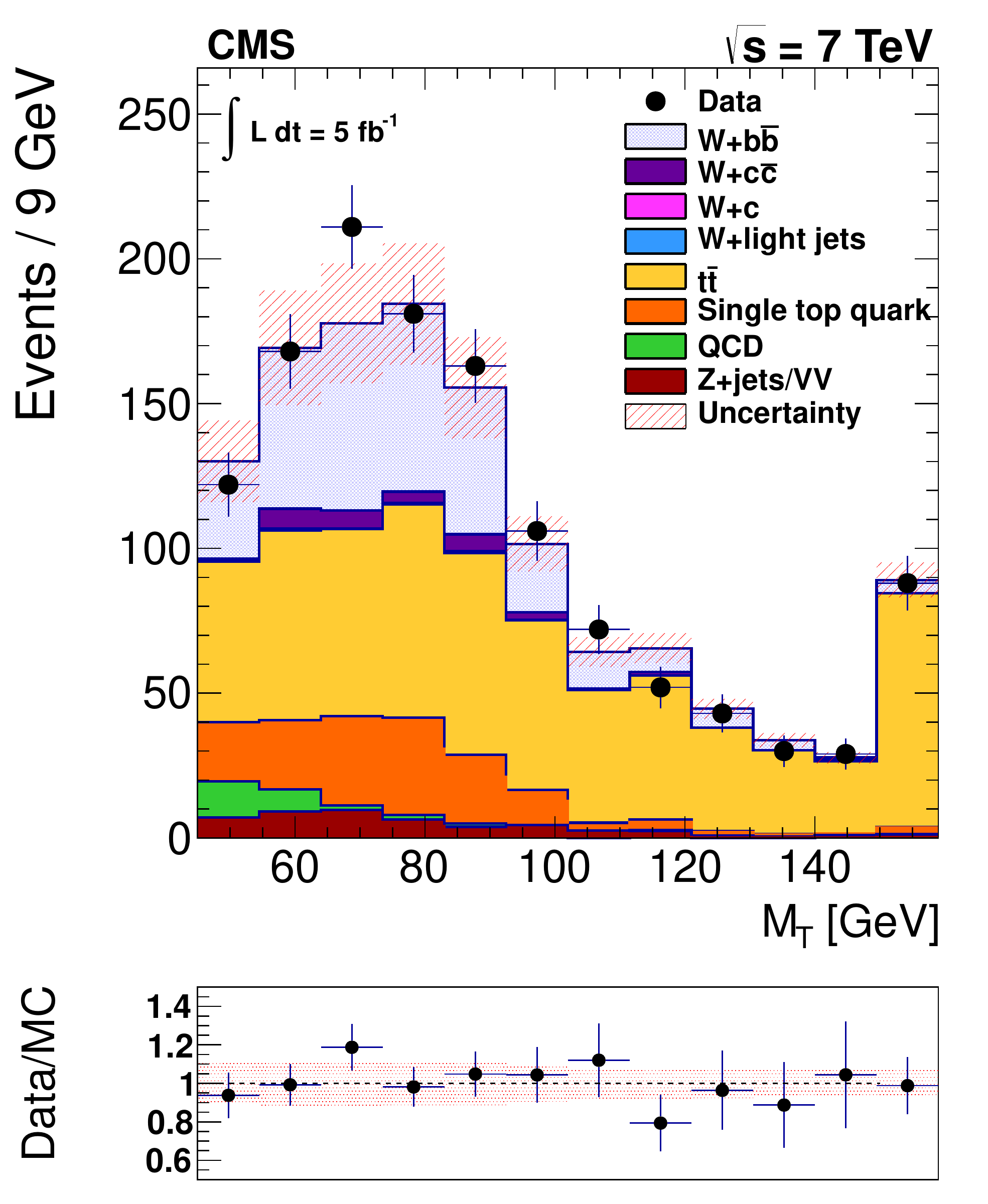}
\caption{The distribution of the angular distance $\Delta R$ between the two selected b jets (\cmsLeft) and
the distribution of the transverse mass $\MT$ of the muon-$\VEtmiss$ system (\cmsRight).
Signal and background yields are taken from
the binned maximum-likelihood fit described in the text.
The uncertainty band corresponds to the uncertainty in the yields as given by the fit.
The last bin in both plots includes overflow events.
The lower panels show the ratio of observed data events
to the total fitted yield. }
\label{figD}
\end{figure}

\begin{figure}[htbp]
\centering
\includegraphics[width=\cmsFigWidth]{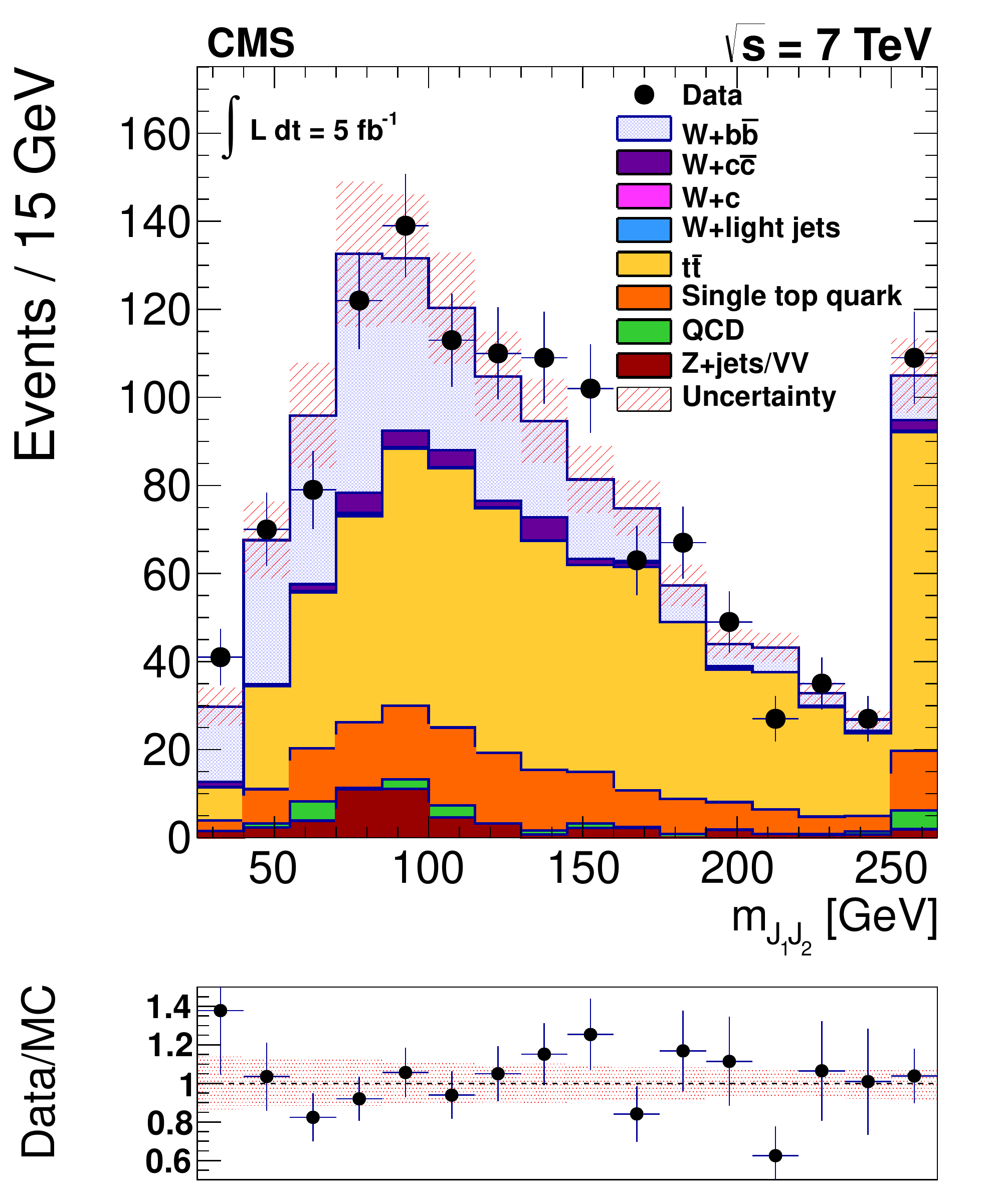}
\includegraphics[width=\cmsFigWidth]{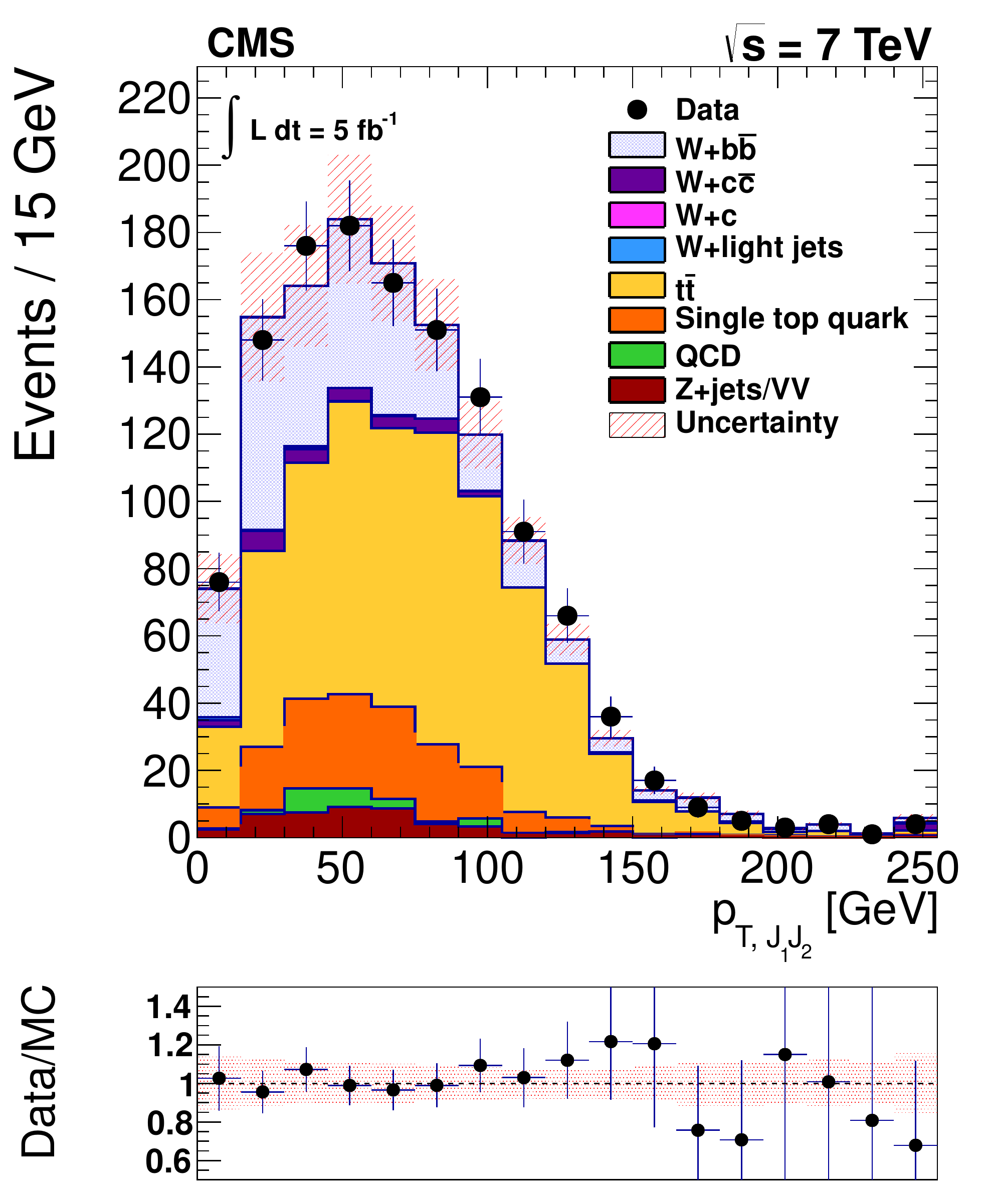}
\caption{The distribution of the invariant mass $m_{\mathrm{J_{1} J_{2}}}$ of the two selected b jets (\cmsLeft) and
the distribution of the transverse momentum of the dijet system, ${\pt}_{,\,\mathrm{J_{1} J_{2}}}$ (\cmsRight).
Signal and background yields are taken from the binned maximum-likelihood fit described in the text.
The uncertainty band corresponds to the total uncertainty in the fitted yields.
The last bin in both figures includes overflow events.  The lower panels show the ratio of observed data events
to the total fitted yield. }
\label{figE}
\end{figure}

In addition to this measurement of the production cross section, we have explored the kinematics of the $\Wbb$ system.
The angular distance between the two selected b jets, $\Delta R_{\mathrm{J_1,J_2}}=\sqrt{\smash[b]{(\Delta \eta_{\mathrm{J_1,J_2}})^2+(\Delta \phi_{\mathrm{J_1,J_2}})}^2}$,
is compared to the SM prediction in Fig.~\ref{figD} (\cmsLeft).
Signal and background yields are taken from
the binned maximum-likelihood fit, and their shapes from Monte Carlo simulations or data as described in Section~\ref{sectionEventSelection}.
The minimum separation of 0.5  between the two jets is  an important aspect of the phase space definition, as discussed in the introduction.
Figure~\ref{figD} (\cmsRight) compares the $\MT$ distribution to the SM predictions.
Figure~\ref{figE} shows the invariant mass of the two
selected b-quark jets ($m_{\mathrm{J_{1} J_{2}}}$) as well as the transverse momentum of the system formed by
the two b-quark jets (${\pt}_{,\,\mathrm{J_{1} J_{2}}}$).
The simulation describes the observed distributions well.

\section{Summary}

In summary, we have presented a measurement of the $\Wbb$ production
cross section in proton-proton collisions at 7\TeV. The $\PW+\bbbar$ events
have been  selected in the $\PW \to \mu\nu$ decay mode with a
muon of $\pt>25\GeV$ and $\abs{\eta}<2.1$, and two b jets of $\pt>25\GeV$ and $\abs{\eta}<2.4$.
The data sample corresponds to an integrated luminosity of $5.0\fbinv$.
The measured fiducial cross section for production of a $\PW$ boson and  two b jets,
$\sigma(\mathrm{pp}\to \PW + \bbbar)\times {\mathcal{B}}(\Wmn) =0.53\pm  0.05\stat \pm 0.09 \syst \pm 0.06 \theo \pm 0.01\lum \unit{pb}$,
is in agreement with
the SM prediction
of $0.55 \pm0.03\MCFMerr \pm 0.01\haderr  \pm 0.05\DPSerr\unit{pb}$, which accounts for double-parton scattering production and hadronization effects.

This study provides the first measurement for $\mathrm{pp}\to \Wbb$ production at $7\TeV$ in this particular phase space,
thereby complementing previous measurements performed at the LHC~\cite{Aad:2011kp}, which focused on the production of
$\PW$ bosons accompanied by one identified b jet.
The precision of the measured cross section approaches that of theoretical predictions at NNLO,
thus enabling sensitive tests of perturbative calculations in the SM.

\section*{Acknowledgements}

We congratulate our colleagues in the CERN accelerator departments for the excellent performance of the LHC and thank the technical and administrative staffs at CERN and at other CMS institutes for their contributions to the success of the CMS effort. In addition, we gratefully acknowledge the computing centers and personnel of the Worldwide LHC Computing Grid for delivering so effectively the computing infrastructure essential to our analyses. Finally, we acknowledge the enduring support for the construction and operation of the LHC and the CMS detector provided by the following funding agencies: BMWF and FWF (Austria); FNRS and FWO (Belgium); CNPq, CAPES, FAPERJ, and FAPESP (Brazil); MES (Bulgaria); CERN; CAS, MoST, and NSFC (China); COLCIENCIAS (Colombia); MSES and CSF (Croatia); RPF (Cyprus); MoER, SF0690030s09 and ERDF (Estonia); Academy of Finland, MEC, and HIP (Finland); CEA and CNRS/IN2P3 (France); BMBF, DFG, and HGF (Germany); GSRT (Greece); OTKA and NIH (Hungary); DAE and DST (India); IPM (Iran); SFI (Ireland); INFN (Italy); NRF and WCU (Republic of Korea); LAS (Lithuania); CINVESTAV, CONACYT, SEP, and UASLP-FAI (Mexico); MBIE (New Zealand); PAEC (Pakistan); MSHE and NSC (Poland); FCT (Portugal); JINR (Dubna); MON, RosAtom, RAS and RFBR (Russia); MESTD (Serbia); SEIDI and CPAN (Spain); Swiss Funding Agencies (Switzerland); NSC (Taipei); ThEPCenter, IPST, STAR and NSTDA (Thailand); TUBITAK and TAEK (Turkey); NASU (Ukraine); STFC (United Kingdom); DOE and NSF (USA).

Individuals have received support from the Marie-Curie programme and the European Research Council and EPLANET (European Union); the Leventis Foundation; the A. P. Sloan Foundation; the Alexander von Humboldt Foundation; the Belgian Federal Science Policy Office; the Fonds pour la Formation \`a la Recherche dans l'Industrie et dans l'Agriculture (FRIA-Belgium); the Agentschap voor Innovatie door Wetenschap en Technologie (IWT-Belgium); the Ministry of Education, Youth and Sports (MEYS) of Czech Republic; the Council of Science and Industrial Research, India; the Compagnia di San Paolo (Torino); the HOMING PLUS programme of Foundation for Polish Science, cofinanced by EU, Regional Development Fund; and the Thalis and Aristeia programmes cofinanced by EU-ESF and the Greek NSRF.

\clearpage
\bibliography{auto_generated}

\cleardoublepage \appendix\section{The CMS Collaboration \label{app:collab}}\begin{sloppypar}\hyphenpenalty=5000\widowpenalty=500\clubpenalty=5000\textbf{Yerevan Physics Institute,  Yerevan,  Armenia}\\*[0pt]
S.~Chatrchyan, V.~Khachatryan, A.M.~Sirunyan, A.~Tumasyan
\vskip\cmsinstskip
\textbf{Institut f\"{u}r Hochenergiephysik der OeAW,  Wien,  Austria}\\*[0pt]
W.~Adam, T.~Bergauer, M.~Dragicevic, J.~Er\"{o}, C.~Fabjan\cmsAuthorMark{1}, M.~Friedl, R.~Fr\"{u}hwirth\cmsAuthorMark{1}, V.M.~Ghete, N.~H\"{o}rmann, J.~Hrubec, M.~Jeitler\cmsAuthorMark{1}, W.~Kiesenhofer, V.~Kn\"{u}nz, M.~Krammer\cmsAuthorMark{1}, I.~Kr\"{a}tschmer, D.~Liko, I.~Mikulec, D.~Rabady\cmsAuthorMark{2}, B.~Rahbaran, H.~Rohringer, R.~Sch\"{o}fbeck, J.~Strauss, A.~Taurok, W.~Treberer-Treberspurg, W.~Waltenberger, C.-E.~Wulz\cmsAuthorMark{1}
% ...
\vskip\cmsinstskip
% ...
1:~~Also at Vienna University of Technology, Vienna, Austria\\
2:~~Also at CERN, European Organization for Nuclear Research, Geneva, Switzerland\\
% ...

\end{sloppypar}
\end{document}